\documentclass[prl,tightenlines, aps]{revtex4}
\usepackage{graphicx}
\usepackage{float}
\begin{document}

\title {Equilibration in an Interacting Field Theory}

\author{ E. Vaz${}^{a}$, M.E. Carrington${}^{b,c}$,  R.
Kobes${}^{a,c}$, and G. Kunstatter${}^{a,c}$.}

 \affiliation{ ${}^a$ Department of Physics, University of Winnipeg, Winnipeg, 
Manitoba, R3B 2E9 Canada \\
${}^b$ Department of Physics and Astronomy, Brandon University, Brandon, 
Manitoba,
R7A 6A9 Canada\\
 ${}^c$  Winnipeg Institute for Theoretical Physics, Winnipeg, Manitoba }


\begin{abstract}
 We use a combination of perturbation theory and numerical techniques to
study the equilibration of two interacting fields which are initially
at thermal equilibrium at different temperatures. Using
standard rules of quantum field theory, we examine the 
 master equations that describe the time evolution of the distribution 
functions for the two coupled systems. By making a few reasonable assumptions
we reduce the resulting coupled integral/differential equations to a pair of
differential equations that can be solved numerically. Our results
show with good accuracy how these coupled systems approach a common
 equilibrium temperature.

\end{abstract}

\maketitle

Non-equilibrium physics is a topic of increasing interest in physics today.  
Such effects are expected to play an important role in many 
contexts, such as the formation of the quark gluon plasma \cite{hvi} and 
electroweak baryogenesis \cite{bg}. Compared to the equilibrium case, 
the details of non--equilibrium processes are difficult 
to analyze,
and a variety of different
 approaches have
been used: analysis of the equations of motion \cite{devega},
renormalization group techniques \cite{boyanovsky}, 
kinetic \cite{kinetic} and transport \cite{transport} theory, and 
thermo field dynamics \cite{tfd}. 
In this article we develop a new approach, and apply it to a simple 
interacting model
of two fields initially each in thermal equilibrium but at different
temperatures.
We find we can reliably track the  path towards
equilibration of this system using perturbation 
theory applied to the master equation
describing the distribution functions of these fields. The analysis
allows us to obtain quantitative information about this non-equilibrium process
directly from the microscopic field theoretic description of the 
physical system. The method can readily be generalized to 
more complicated, and physically relevant, systems.

The partition function describing the system we are considering is given by,
\begin{equation}
Z=\int d\phi {\rm exp} \{ i\int dx [\frac{1}{2} 
(\partial_\mu \phi_a)^2 - m^2 \phi_a^2 + 
\frac{1}{2} (\partial_\mu \phi_b)^2 - m^2 \phi_b^2 -
\frac{g}{4} |\phi_a \phi_b|^2]\} 
\end{equation}
where $a$ and $b$ are the two field types. We will use the Keldysh 
representation of the closed time path formulation of finite 
temperature field theory \cite{ctp}, so that 
indices $\{1,2\}$ indicate the top/bottom 
branch of the time contour.  

The Schwinger-Dyson equation for the two-point
function for field $a$ has the form \cite{yaffe}, 
\begin{eqnarray}
[(\partial\cdot\partial)_x - (\partial\cdot\partial)_y]S_{12}^a(x,y) 
= \int dz\,&&[S_{11}^a(x,z) \Sigma_{12}^a(z,y) - S_{12}^a(x,z) 
\Sigma_{22}^a(z,y) \nonumber \\
&&- \Sigma_{11}^a(x,z) S_{12}^a(z,y) + 
\Sigma_{12}^a(x,z) S_{22}^a(z,y)] 
\label{sd}
\end{eqnarray}
and similarly for system $b$.
Writing all functions
in terms of an average $X=\frac{1}{2}(x+y)$ and a relative
$s=x-y\,;~~s'= x-z$ coordinate,
and dropping terms $(s-s')$ and $s'$ relative to $X$, taking the  
transform $\int ds\,e^{-iks}$ of (\ref{sd}) yields
\begin{equation}
\label{1st}
-2ik_\mu \frac{\partial}{\partial X_\mu} S^a_{12}(X,k) = 
S^a_{21}(X,k)\Sigma^a_{12}(X,k) - \Sigma^a_{21}(X,k)S^a_{12}(X,k)
\end{equation}
\par
We now make the ansatz for the propagators,
\begin{eqnarray}
&& S^a_{12}(X,k) = -\frac{i\pi}{\omega^a_k}
[\delta(k_0-\omega^a_k) n^a(k_0,X) + \delta(k_0+\omega^a_k) 
(1+n^a(-k_0,X))] \nonumber \\
&&  S^a_{21}(X,k) = -\frac{i\pi}{\omega^a_k}
[\delta(k_0-\omega^a_k)(1+ n^a(k_0,X)) + \delta(k_0+\omega^a_k) 
n^a(-k_0,X)] \nonumber 
\end{eqnarray}
where $\omega_a(k) = \sqrt{k^2+m_a^2}$. Assuming now
spatial homogeneity, 
equating the positive frequency parts in (\ref{1st}) gives
\begin{equation}
\frac{d n_a (\omega^a_k,t)}{dt} = -n_a(\omega^a_k,t) 
\Gamma_d(\omega^a_k) + [1+ n_a(\omega_k^a,t)] \Gamma_i(\omega_k^a) 
\label{difeq}
\end{equation}
where  $n^a(\omega_k^a,X) \equiv n^a_k$ and
\begin{equation}
\Gamma_i(\omega_k^a) = \frac{i}{2\omega_k^a} 
\Sigma_{12}^a \,;~~~~\Gamma_d(\omega_k^a) = 
\frac{i}{2\omega_k^a}\Sigma_{21}^a
\label{rates}
\end{equation}
are, respectively, rates associated with processes which increase/decrease
the number of particles present of that species.
These rates are complicated functionals
of the distribution functions, and so (\ref{difeq}) is very difficult to
solve in general. However, matters become simpler in the case that each  
distribution function is, at a given time, close to
a thermal equilibrium distribution, as
we will now describe.
\par
We assume that the two fields start out in
(different) thermal equilibrium states with initial conditions
\begin{equation}
n_x(k_0,0) = \frac{1}{e^{k_0 \beta_x} -1}\,;~~~~x=a,b
\label{ics}
\end{equation}
We further assume  that the interactions
are described by the two-loop sunset self-energy. For given thermal
equilibrium distribution functions, a detailed calculation reveals that:
\begin{eqnarray}
\Sigma^a_{21}(k_0,k) = -\frac{i}{16\pi^2} &&\int^\infty _0 
dp\,\frac{p^2}{\omega_p^a}\int ^1_{-1} dx\, \nonumber \\
&&[(1+ n(\omega_p^a \beta_a)) {\cal L}^b(k_0,\vec k,\omega_p^a,\vec p)
 + n(\omega_p^a\beta_a) {\cal L}^b(k_0,\vec k,-\omega_p^a,\vec p) ] 
\label{21}
\end{eqnarray}
\begin{eqnarray}
\Sigma^a_{12}(k_0,k) = -\frac{i}{16\pi^2} &&\int^\infty _0 
dp\,\frac{p^2}{\omega_p^a}\int ^1_{-1} dx\, \nonumber \\
&&[ n(\omega_p^a\beta_a){\cal L}^b(\omega_p^a,\vec p,k_0,\vec k) 
+ (1+ n(\omega_p^a \beta_a)){\cal L}^b(-\omega_p^a,\vec p,k_0,\vec k)]
\label{12}
\end{eqnarray}
where $\omega_x(k) = \sqrt{k^2+m_x^2}$ and
the function ${\cal L}^b(K,P)$ is given by,
\begin{eqnarray}
 {\cal L}^b(K,P)&=& \frac{ g^2
  [1+n^b\{({k}^0-{p}^0)\beta_b\}]}{
    8\pi
  |{\vec k}{-}{\vec p}|\beta_b}\times
 \left\{\,
 \theta( -({K}{-}P)^2 )
 \ln
 \left|\frac
 {1 - \exp(- r_{+}\beta_b)}{
     1 - \exp(- r_{-}\beta_b)}\right|
 \right.
 \nonumber\\
 \noalign{\smallskip}
 \; & \displaystyle \; & \displaystyle
 \qquad {}
 +\left.
 \theta( ({K}{-}{P})^2{-}4 m_b^2 )
 \ln
 \left|
 \frac{
  \sinh( r_{+}\beta_b/2)
        }{
  \sinh( r_{-}\beta_b/2)
 }
 \right|\,
 \right\}
 \;,
 \label{rung}
 \end{eqnarray}
 where
 \begin{eqnarray}
 r_{\pm} &\equiv&
 {1\over 2}
     \left(
        |{\vec k}{-}{\vec p}|
        \sqrt{1{-}4 m_b^2/ ({K}{-}{P})^2}
        \pm ({k}^0{-}{p}^0)
    \right), \nonumber\\
|{\vec k}{-}{\vec p}|&=&\sqrt{k^2+p^2-2kp x}, \qquad\quad
 P^2 = p_0^2 - p^2.
\end{eqnarray}
It is important to note that for the initial conditions (\ref{ics}), 
the KMS condition
\begin{eqnarray}
\Gamma_d = e^{\omega_k^a \beta_a}\Gamma_i,
\label{kms}
\end{eqnarray}
is satisfied if and only if $\beta_a=\beta_b$; i.e. the combined system
is in thermal equilibrium.
\par
We take the following approach to solving (\ref{difeq}). We begin
with initial conditions (\ref{ics}) which describe the
two fields at two different (inverse) temperatures $\beta_a$ and
$\beta_b$.
We assume that the equilibration time for each separate system is very 
small compared to the time scale of equilibration for the
coupled system, so that the distribution function for each system remains
close to a thermal one as the systems evolve. Now,
near $t=0$, the
distribution functions which appear within the rates in (\ref{rates}) 
can be replaced by their thermal
values, and (\ref{difeq}) has the solution
\begin{equation}
n_a(\omega_k^a,t) = \frac{\Gamma_i}{\Gamma_d - \Gamma_i} 
+ C(\omega_k^a){\rm exp}(-[\Gamma_d - \Gamma_i] t)
\label{soln}
\end{equation}
with $C(\omega_k^a)$ determined from the initial condition.
Note that if $\beta_a=\beta_b$, the KMS condition (\ref{kms}) is satisfied, and
substituting into (\ref{soln}) we find $C(\omega_k^a)=0$, so that  
$n_a(\omega_k^a,t) = n_a(\omega_k^a,0)$; this simply
means that the combined system stays at equilibrium, as expected.
Thus, for a suitably short time interval 
$ \delta t \ll 1/[\Gamma_d - \Gamma_i]$, (\ref{soln}) gives the approximate
time evolution of each distribution function.
To integrate (\ref{difeq}) to finite time intervals, we first let 
the systems evolve according to (\ref{soln}) for a time $\delta t$,
and the subsystems are assumed to come
separately to thermal equilibrium. We then fit the 
distribution functions resulting from (\ref{soln})
to equilibrium distribution functions of the form:
\begin{eqnarray}
n_x(k_0, |{\vec{k}}|) = \frac{1}{e^{\beta_x\sqrt{k^2+m_x^2}}-1}
\end{eqnarray}
by adjusting both the parameters characterizing the
inverse temperature $\beta_x$ and a (temperature dependent)
mass $m_x$ for each subsystem. 
The thermal distributions resulting from this best--fit procedure 
are then used as the initial conditions for the
next time iteration. This  results in a second evolved set of
distribution functions  which again are  fit to thermal
distributions. This process is continued until
(hopefully) the best fit temperatures and masses, and hence
the new distribution functions, change little
from one time step to the next.
\par
\begin{figure}[H]
\begin{center}
\includegraphics[height=5cm]{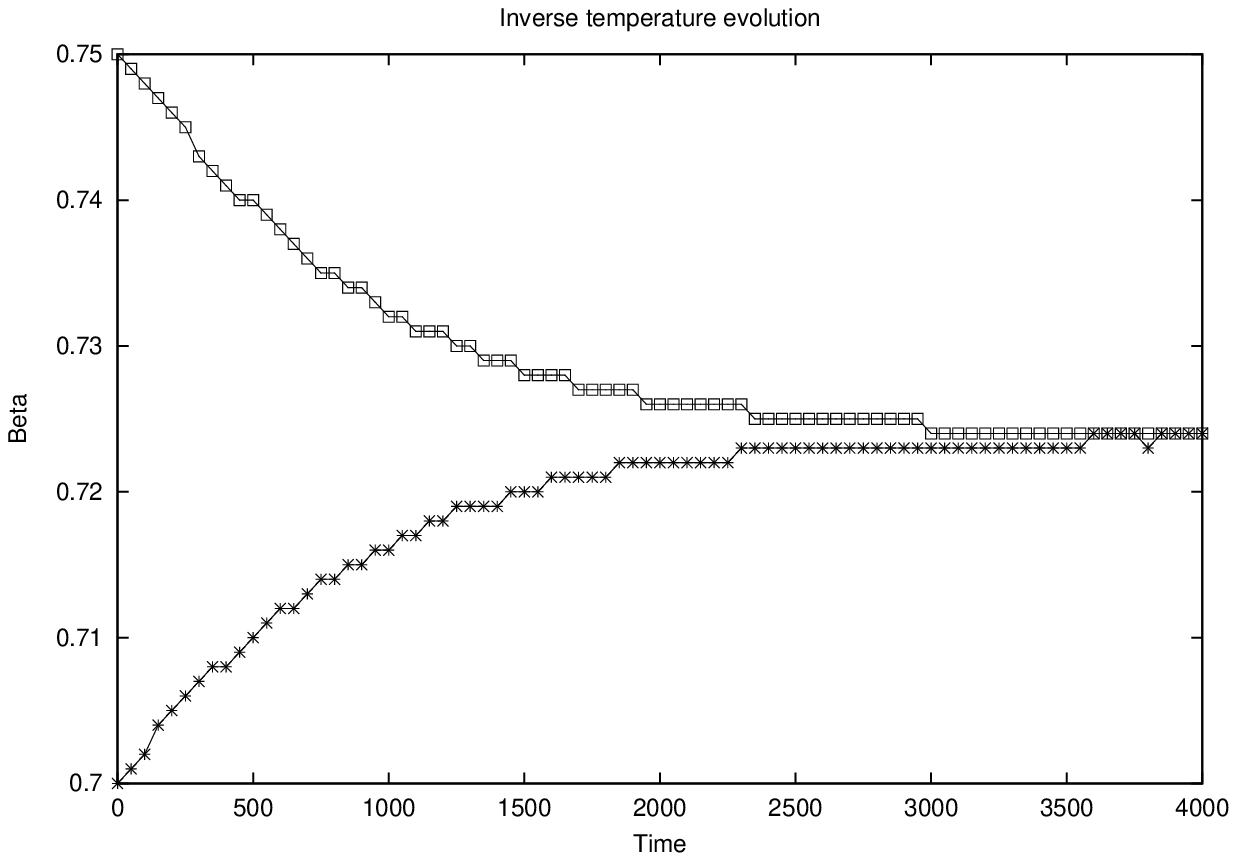}
\includegraphics[height=5cm]{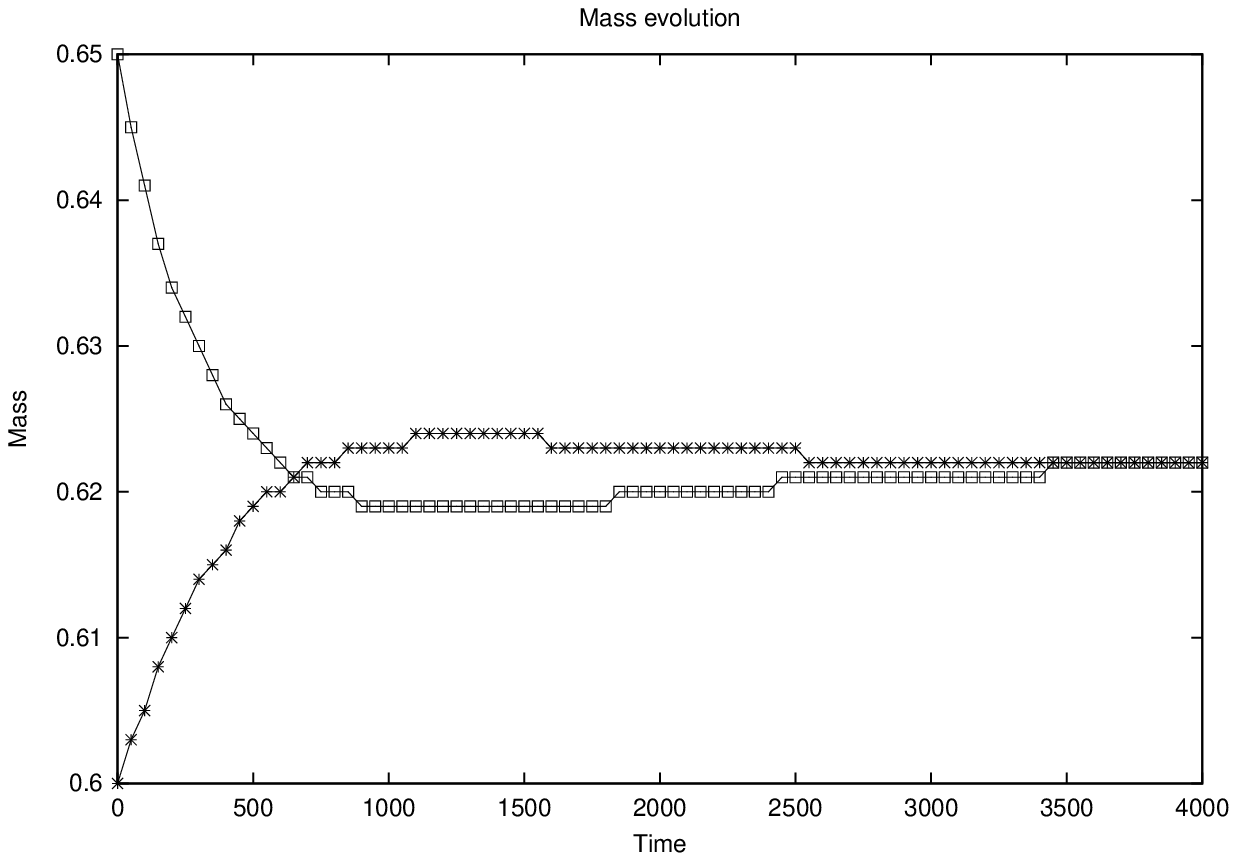}
\end{center}
\caption{Evolution of the inverse temperature (left) and the mass (right). 
The crossed points
represent the $a$ field, and  the boxed points represent the $b$ field.}
\label{massbeta}
\end{figure}

We present some results of this procedure for initial
temperatures of 0.7 and 0.75 and initial masses of
0.6 and 0.65 for, respectively, the $a$ and $b$ fields.
The resulting evolution, for time steps of 50 units, for
the inverse temperature and masses appears Fig.~(\ref{massbeta}) --
note that in both cases a common equilibrium is reached.
\par
\par
\begin{figure}[H]
\begin{center}
\includegraphics[height=5cm]{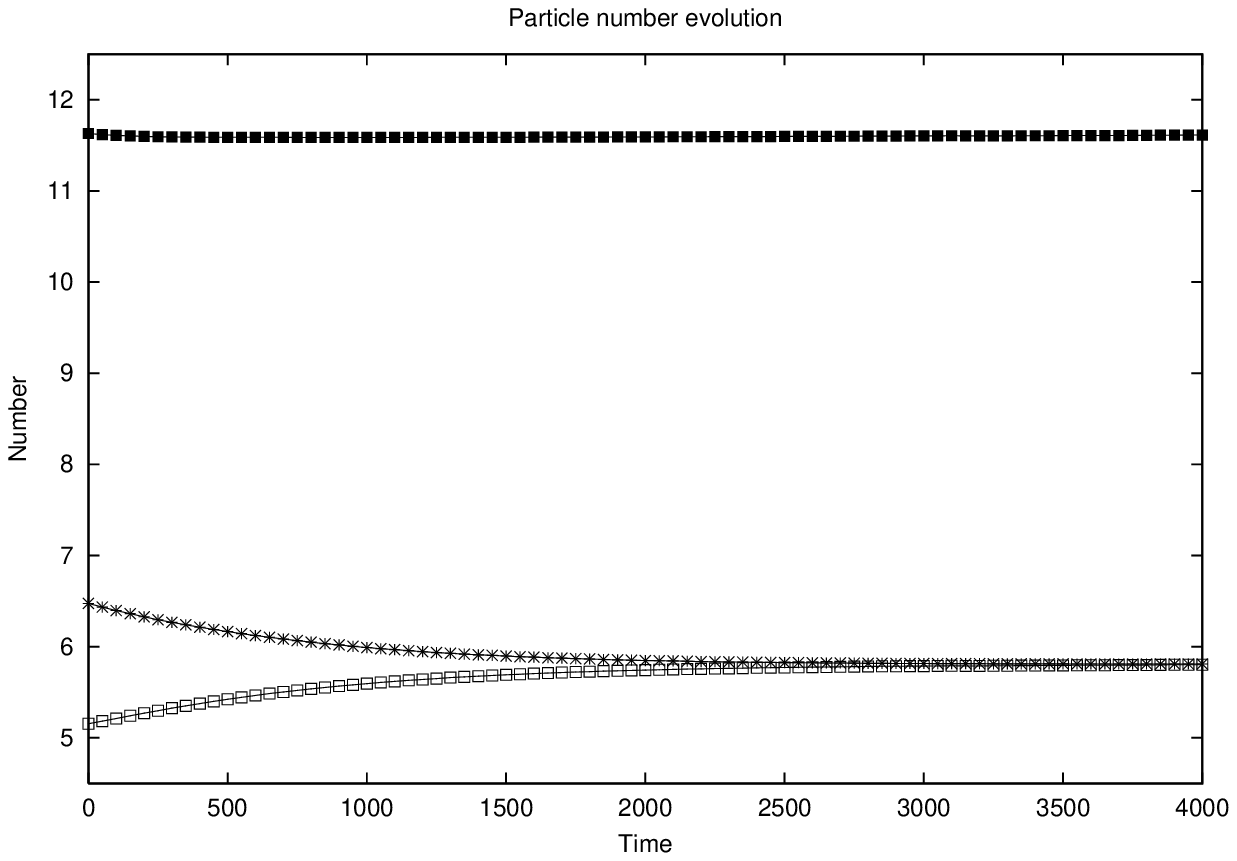}
\includegraphics[height=5cm]{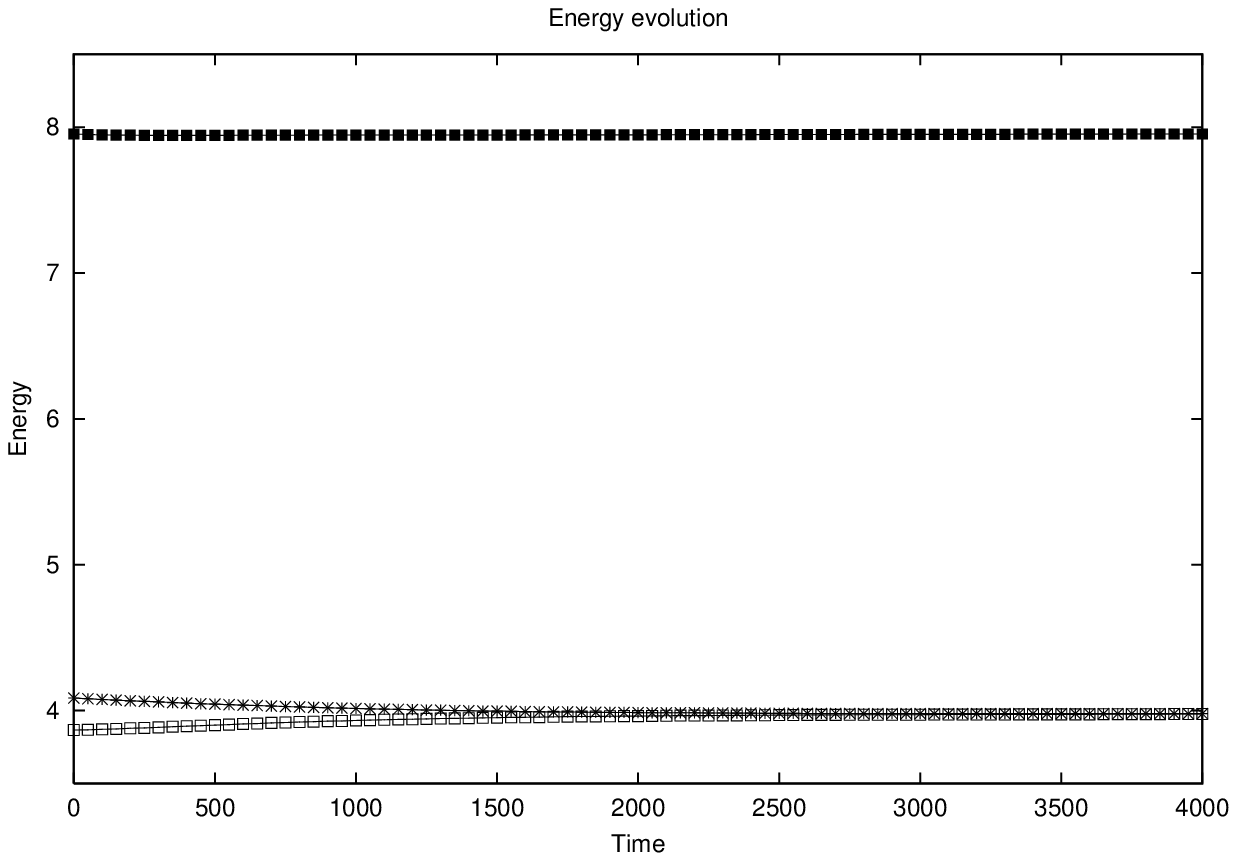}
\end{center}
\caption{Evolution of the particle number (left) and the energy (right). 
The crossed points
represent the $a$ field, the boxed points represent the $b$ field, and 
the triangulated points of the top
line represent the sum.}
\label{numberenergy}
\end{figure}

Using these distributions we can follow, for each species, 
the evolution of the particle number $N(t)$ and energy $E(t)$:
\begin{eqnarray}
N(t) &=& \int_0^\infty k^2\,dk \ n(k,\beta, m)\nonumber\\
E(t) &=& \frac{1}{N(t)}
\int_0^\infty k^2\,dk \ \sqrt{k^2+m^2}\ n(k,\beta, m)
\end{eqnarray}
Again for time steps of 50 units, these results appear in
Fig.~(\ref{numberenergy}) -- to a good 
approximation the total
particle number and the total energy remains constant, although
small numerical deviations are observed.
\par
\par
The preceding results were obtained using a basic time step of
50 units.
We can check the consistency of the results by following the
evolution of the system for different basic time steps and then comparing
the results at common times. Results are shown in the following
table for the evolution of $\beta$ for field $a$ using basic time 
steps of 50, 100, 200, 400, and 800 units.
\begin{table}[H]
\begin{center}
\begin{tabular}{||l|l|l|l|l|l||l||l|l|l|l|l|l||}
\hline
Time & 50 & 100 & 200 & 400 & 800 &$\qquad\qquad$ &  Time &
50 & 100 & 200 & 400 & 800\\ \hline\hline
0 & 0.700 & 0.700 & 0.700 & 0.700 & 0.700 & &
550 & 0.711 &  &  &  &   \\ \hline 
50 & 0.701 &  &  &  &   & &
600 & 0.712 & 0.712 & 0.712 &  &   \\ \hline 
100 & 0.702 & 0.702 &  &  &   & &
650 & 0.712 &  &  &  &    \\ \hline 
150 & 0.704 &  &  &  &   & &
700 & 0.713 & 0.713 &  &  &   \\ \hline 
200 & 0.705 & 0.705 & 0.705 &  &   & &
750 & 0.714 &  &  &  &   \\ \hline 
250 & 0.706 &  &  &  &   & &
800 & 0.714 & 0.714 & 0.715 & 0.715 & 0.716 \\ \hline 
300 & 0.707 & 0.707 &  &  &   & &
1000 & 0.716 & 0.717 & 0.717 &  &     \\ \hline 
350 & 0.708 &  &  &  &  & &
1200 & 0.718 & 0.718 & 0.719 & 0.719 &    \\ \hline 
400 & 0.708 & 0.709 & 0.709 & 0.709 &  
 & & 1600 & 0.721 & 0.721 & 0.721 & 0.722 & 0.724 \\ \hline 
450 & 0.709 &  &  &  &  
 & & 2000 & 0.722 & 0.722 & 0.722 & 0.723 &  \\ \hline 
500 & 0.710 & 0.710 &  &  &  
 & & 2400 & 0.723 & 0.723 & 0.723 & 0.724 & 0.725 \\ \hline 
\hline
\end{tabular}
\end{center}
\caption{Comparison of the evolution of the inverse temperature of 
$a$ for basic time steps of 50, 100, 200, 400, and 800 units.}
\end{table}
We see that the results amongst the different time steps
agree to a good approximation at the corresponding times. However,
as is expected, for larger time steps some disagreement starts
to emerge. 

In conclusion, we have shown that equilibrium is achieved in the simple 
model we have considered.  Starting from two separate systems at 
temperatures that are close enough together so that the combined system is 
never very far from equilibrium, we fit to a series of equilibrium 
distributions using time steps that are small compared with the 
equilibration time. The result is that the combined system reaches 
equilibrium at a temperature between the temperatures of the two 
original systems.  In addition, we have shown that both particle number 
and energy are conserved.  It would be interesting to try to relax the 
assumption that the initial temperatures of the two initial systems are 
close together.  Allowing for a larger difference in initial temperatures 
would create a more highly non-equilibrium state in the early stages of 
the evolution.  Such a situation might be handled by using a more 
complex ansatz for the distribution functions of the evolving system.

\end{document}